\documentclass[10pt]{iopart}

\usepackage{iopams}

\def\de#1/de#2{\frac{\partial {#1}}{\partial {#2}}}
\def\De#1/de#2{\dfrac{\partial {#1}}{\partial {#2}}}

\begin{document}

\title[A comment on ``The Cauchy problem of $f(R)$ gravity'']{A comment on ``The Cauchy problem of $f(R)$
gravity'', {\it Class. Quantum Grav.}, {\bf 24}, 5667 (2007)}

\author{S. Capozziello$^1$, S. Vignolo$^2$}
\address{$^1$Dipartimento di Scienze Fisiche and INFN, Sez. di Napoli, Universit\'a di Napoli ``Federico II'',
Compl. Univ. di Monte S. Angelo, Ed. G., Via Cinthia, I-80126
Napoli, Italy}\ead{capozziello@na.infn.it}

\address{$^2$DIPTEM Sez. Metodi e Modelli Matematici, Universit\`a di Genova
Piazzale Kennedy, Pad. D - 16129 Genova,
Italy}\ead{vignolo@diptem.unige.it}

\begin{abstract}
A critical comment on [N. Lanahan--Tremblay and V. Faraoni, 2007,
{\it Class. Quantum Grav.}, {\bf 24}, 5667] is given discussing
the well-formulation of the Chauchy problem for $f(R)$-gravity in
metric-affine theories.
\end{abstract}


\bigskip
In Ref. \cite{Faraoni}, the authors question the viability of
$f(R)$-theories of gravity in the Palatini metric-affine
formulation, since the Cauchy problem, following their approach,
cannot be well--formulated. From their viewpoint, this shortcoming
is present also in vacuo. Such a conclusion can be debated and, on
the basis of the results which we are going to present in this
Note, the well--formulation of the initial value problem can be
demonstrated and  no objections can be made on the viability of
metric--affine $f(R)$-gravity. On the other hand, the
well--posedness of the Cauchy problem has to be investigated.

Firstly, we recall that, in the metric-affine formulation of
$f(R)$-gravity, the dynamical fields are given by the couple of
functions $(g,\Gamma)\/$ where $g\/$ is the metric and  $\Gamma\/$
is the linear connection. In vacuo, the field equations are
obtained by varying the following action with respect to the
metric and the connection

\begin{equation}\label{2.0}
{\cal A}\/(g,\Gamma)=\int{\sqrt{|g|}f\/(R)\,ds}
\end{equation}
where $f(R)$ is a real function, $R\/(g,\Gamma) = g^{ij}R_{ij}\/$
(with $R_{ij}:= R^h_{\;\;ihj}\/$) is the scalar curvature
associated to the dynamical connection $\Gamma\/$.

More precisely, in the approach with torsion, one can ask for a
metric connection $\Gamma$ with torsion different from zero while,
in the Palatini approach, the $\Gamma$ is non-metric but torsion
is null \cite{CCSV1}.

In vacuo, the field equations for $f(R)$-gravity with torsion are
\cite{CCSV1}
\begin{equation}\label{2.1a}
f'\/(R)R_{ij} - \frac{1}{2}f\/(R)g_{ij}=0\,,
\end{equation}
\begin{equation}\label{2.1b}
T_{ij}^{\;\;\;h} = -
\frac{1}{2f'}\de{f'}/de{x^p}\/\left(\delta^p_i\delta^h_j -
\delta^p_j\delta^h_i\right)\,,
\end{equation}
while the field equations for $f(R)$-gravity  {\it \`{a} la}
Palatini are \cite{francaviglia1,francaviglia2,Sotiriou,Sotiriou-Liberati1,Olmo}
\begin{equation}\label{2.2a}
f'\/(R)R_{ij} - \frac{1}{2}f\/(R)g_{ij}=0\,,
\end{equation}
\begin{equation}\label{2.2b}
\nabla_k\/(f'(R)g_{ij})=0\,.
\end{equation}
In both cases, considering the trace of  Einstein-like field
Eqs.(\ref{2.1a}) and (\ref{2.2a}), one gets
\begin{equation}\label{2.3}
f'\/(R)R  - 2f\/(R)=0\,.
\end{equation}
It is easy to conclude that the scalar curvature $R$ is a constant
coinciding with the solution of Eq. (\ref{2.3}). In this case,
Eqs.(\ref{2.1b}) and (\ref{2.2b}) imply that both dynamical
connections coincide with the  Levi--Civita connection  associated
to the metric $g_{ij}$ which is the solution of the field equations.

In other words, both theories reduce to the Einstein theory plus
cosmological constant. It is well known that General Relativity in
vacuo shows a well--formulated and a well--posed Cauchy problem,
both in the Lagrangian and Hamiltonian ($3+1$ ADM) formulation
\cite{Wald,MTW,yvonne}. This last fact is inconsistent with the
arguments in  \cite{Faraoni}, where the authors state that
Palatini $f(R)$ gravity {\it "has an ill--formulated Cauchy
problem in vacuo and, therefore, can hardly be regarded as a
viable theory"} (see  Sec. 4 of \cite{Faraoni}).

It is worth noticing that metric--affine  $f(R)$-theories of
gravity (in both the Palatini and with torsion approaches) reduce
to the General Relativity with the cosmological constant also for
the coupling with an electromagnetic field (or, more in general,
with Yang--Mills fields). Therefore, also in this case, the Cauchy
problem is well--formulated and well--posed \cite{Wald}. In
general, the same conclusion holds anytime the trace of the
energy--momentum tensor, associated with the matter sources, is
constant since it is straightforward to show that, in such a
circumstance, the theory reduces to the General Relativity with a
cosmological constant \cite{CCSV1}.

Moreover,  in the Lagrangian formulation (covariant approach and
second order time evolution equations), it can be shown that the
Cauchy problem is well--formulated also in the case of coupling
with a perfect fluid or with a Klein--Gordon scalar field
\cite{CV}. This result is achieved by applying the approach
proposed in \cite{Synge} to the Einstein--like field equations
resulting from metric--affine $f(R)$-theories, and developing a
second order analysis (in the time--derivatives). In Ref.
\cite{CV} and in this Note, for {\it well--formulation\/} of the
Cauchy problem, we mean the possibility to solve the field
equations starting from the assigned  initial data on a Cauchy
surface and using suitable constraints for evolution equations.
The adopted notion is consistent with that assumed in
\cite{Faraoni, Salgado}. The only difference is that, in
\cite{Faraoni, Salgado}, a first order analysis has been developed.

In conclusion,  the statements about the non--viability of
metric--affine $f(R)$-gravity, given in \cite{Faraoni}, seem
excessively severe, at least if it is based only on the supposed
ill--formulation of the initial value problem.

In order  to prove whether metric--affine $f(R)$-theories of
gravity are viable or not, the well--posedness of the Cauchy
problem has to be discussed. Namely, the continuous and causal
dependence of the solutions on the initial data has to be proven
(at least in some physically important cases such as the coupling
with  perfect--fluid matter or the coupling with a scalar field).
In our knowledge, the latter is still an open question as well as
an interesting future task.

\end{document}